\documentclass[prl,aps,twocolumn,showpacs]{revtex4}

\usepackage[dvips]{graphicx}
\usepackage{dcolumn}%
\usepackage{amsmath}%
\usepackage{amsfonts}%
\usepackage{amssymb}
 \usepackage[usenames]{color}
\usepackage{appendix}

\providecommand{\U}[1]{\protect\rule{.1in}{.1in}}
\newcommand{\be}{\begin{equation}}
\newcommand{\ee}{\end{equation}}
\newcommand{\bea}{\begin{eqnarray}}
\newcommand{\eea}{\end{eqnarray}}
\newcommand{\bt} {\begin{tabular}}
\newcommand{\et} {\end{tabular}}
\newcommand{\nn}{ \nonumber}
\newcommand{\ds}{\displaystyle}
\newcommand{\ba} {\begin{array}}
\newcommand{\ea} {\end{array}}
\begin{document}

\title{Chiral-induced circularly polarized light emission from a single-molecule junction}

\author{  Natalya A. Zimbovskaya {\footnote{Corresponding author: nzimbovskaya@gmail.com}}}

\affiliation
{Department of Physics and Electronics, University of Puerto Rico-Humacao, CUH Station, Humacao, PR 00791, USA}

\begin{abstract}
In the present work we theoretically analyze electroluminescence  occurring in a biased single-molecule junction with a chiral bridge imitated by a helical chain. We show that optical transitions between electron states of the chiral linker may result in the emission of circular polarized light whose handedness depends on both direction of propagation and the polarity of the bias voltage provided that the coupling between the bridge sites is sufficiently strong. The mechanism controlling this specific light emission does not  depend on the magnetic moments and spin-orbit interactions. It rather relies on the chiral properties of the bridge molecule and on the distribution of the bias voltage between the electrodes in the junction.

\end{abstract}

\date{\today}
\maketitle

\subsection{I. Introduction} 

Chiral optoelectronic materials which can absorb/emit circularly polarized (CP) light presently attract a significant interest. The emerging field of chiral quantum optics \cite{1} focuses on manufacturing and exploiting systems which show chiroptical activity \cite{2,3,4,5}. Chiral coupling between quantum emitters and photons offers notable benefits in various applications \cite{6,7,8,9,10,11,12}.

Presently, several mechanisms responsible for CP light emission are recognized. The most common mechanism is based on the joint effect of electric and magnetic dipole moments (magnetoelectric coupling) \cite{2,3,13,14,15}. The magnetoelectric coupling depends on both  electric and magnetic dipole transitions matrix elements. However, magnetic dipole transitions are significantly weaker than electric ones and this reduces the efficiency of CP light emitting diodes. Also, it was suggested that the CP luminescence may be controlled by spin-orbit interactions \cite{16,17}. To support this hypothesis one may refer to the well establisheed fact that spin-orbit interactions are responsible for the effect of spin-selectivity in the electron transport through chiral molecules \cite{18,19,20,21,22,23,24}.

Nonreciprocal CP light emission, that is emission of the oppositely handed light from the front and the back of a sample is a phenomenon of significant fundamental scientific interest. At the same time, this phenomenon may suggest a valuable opportunity for improving the performance of light emitting diodes. This effect may appear in mechanically deformed thin films. In such materials the handedness of the emitted CP light may depend on the specific orientation of mesoscopic domains emerging in the process of manufacturing of relevant materials \cite{25,26}.

Electrically stimulated nonreciprocal CP luminescence was recently reported in a light-emitting diode based on a chiral polymeric system \cite{12}. The effect was manifested in the presence of electric current flowing in the diode. The handedness of the emitted light appeared to be sensitive to both the current direction and the the light emission direction. An electronic mechanism not related to electromagnetic coupling and/or spin-orbit interactions was proposed to explain the effect. It was suggested that the effect originates from the specific topology of the electron structure which leads to violation of time-reversal symmetry (TRS). As a result, electrons acquire nonequilibrium angular momentum which may be transferred to photons emitted in optical transitions and cause the observed anomalous electroluminescence.

It had been recently shown that coupling of electron orbital motion with light occurring in biased molecular junctions may give rise to the optical angular momentum radiation \cite{27,28,29,30,31,32}. This may happen in junctions where the leads are linked by a chiral molecule and TRS may be broken by the applied bias voltage and the resulting electron current. Thus there exists a close similarity between the mechanism controlling the anomalous CP light emission in polymeric systems reported in Ref.\cite{12} and that responsible for angular momentum radiation from chiral molecular bridges in single molecule junctions \cite{32}. One may expect a nonreciprocal CP luminescence to appear in such systems unlike the luminescence repeatedly observed in common molecules \cite{33,34,35,36,37,38}.

The present work is inspired by the above results. Here, we theoretically analyze  CP light emission in a current carrying single molecule junction with a chiral linker. Following Ref.\cite{32} we simulate the linker with a helical chain, as displayed in Fig.1. We show that the chain geometry which causes the optical angular momentum radiation may also be responsible for the nonreciprocal CP luminescence.

\subsection{II.Main equations}

In the following analysis we use a tight-binding Hamiltonian to describe the molecular bridge. The Peierls substitution \cite{39} with the phase factor $\theta_{mn}$ is employed to account for the interaction of the molecule with radiation fields. Assuming that each site on the molecular bridge chain is assigned an on-site energy $E_0=0$, the Hamiltonian has the form:
\be
H_{mol}=\sum_{m,n}H_{mn}c^{\dagger}_{m}c_n\exp(i\theta_{mn})                  \label{1}
\ee
where$H_{mn}$ ($m\neq n$) are matrix elements between the sites $m$ and $n$ and $c^{\dagger}_{m}$ and $c_n$ are the creation and annihilation operators at the corresponding sites. The phase factor $\theta_{mn}$ may be computed by integration of the vector potential of the radiation field ${\bf A}({\bf r})$ (which obeys the gauge $\nabla\cdot{\bf A}=0$) along the line connecting the sites located at the positions ${\bf r}_m$ and ${\bf r}_n$:
\be
\theta_{mn}=\int_{{\bf r}_m}^{{\bf r}_n} {\bf A}\cdot d{\bf l}                  \label{2}
\ee
\begin{figure}[t] 
\begin{center}
 \includegraphics[width=4.2 cm,height=3.1cm]{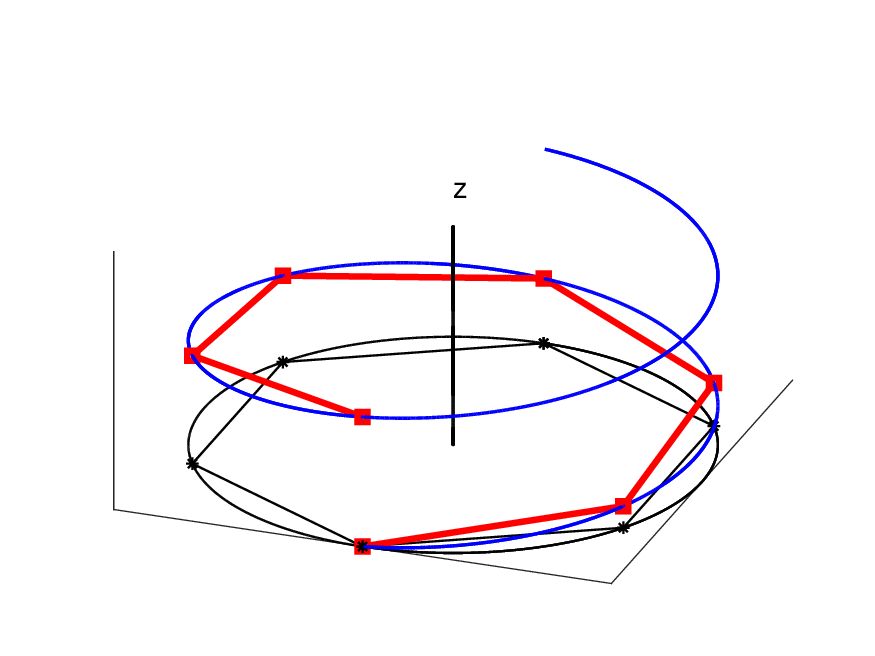}
\includegraphics[width=4.2 cm,height=3.1cm]{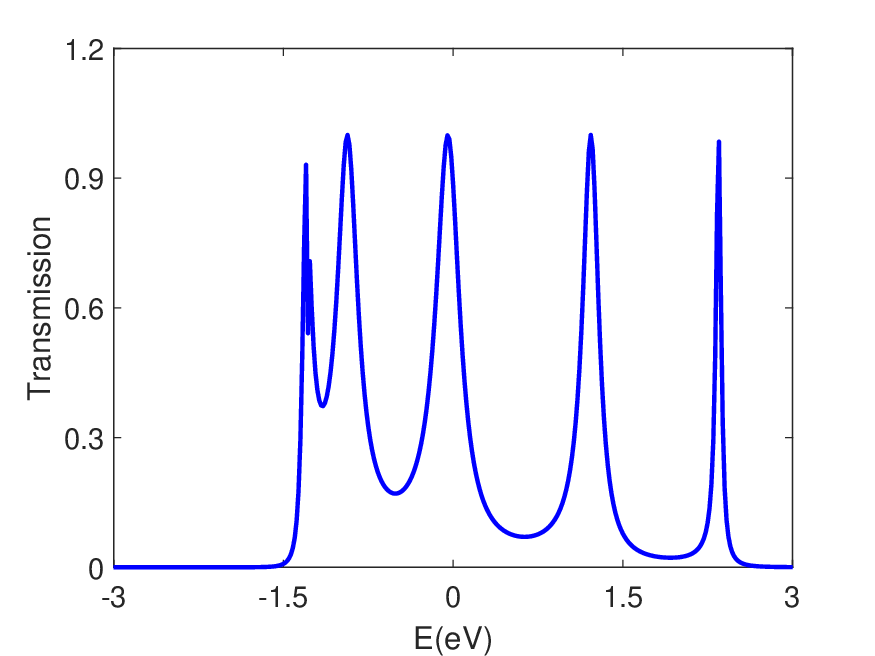}
\caption{Schematics of a helical bridge including six sites (left) and the electron transmission plotted for a six site helical bridge at $E_0=0$, $\Gamma_L=\Gamma_R=0.2$eV, $t=1$eV, $\gamma=0.4$eV and the radius of gyration of the helix $r_g=0.5$nm (right).
}
\label{rateI}
\end{center}\end{figure}
Assuming that the spacings ${\bf r}_n-{\bf r}_m$ are small, it is possible to expand the exponential in Eq.(\ref{1}) in power series keeping the terms linear in ${\bf A}$. Then the bridge Hamiltonian may be presented as a sum:
\be
H_{mol}=\sum_{m,n}\left(H_{mn}+\sum_{k,\mu} M^{k\mu}_{mn}A_{\mu}({\bf r}_k)\right)c^{\dagger}_{m}c_n               \label{3}
\ee
with $M^{k\mu}_{mn}=\ds\frac{ie}{\hbar}H_{mn}({\bf r}_m-{\bf r}_n)(\delta_{km}+\delta_{kn})$ and $\mu=\{x,y,z\}$. Further we choose the coordinate system where $z$ axis is aligned with the axis of the helical molecular bridge, as shown in Fig.1 

Adopting the Hamiltonian given by Eq.(\ref{3}) we ignore spin degrees of freedom for electrons on the bridge. This is a sound simplification highlighting that we consider orbital motion of electrons as a sole reason for the CP luminescence using the same approach as was employed in earlier works \cite{12,31,32}. Also, we follow Ref.\cite{32} and assume that the molecule does not affect radiation fields and the vector potential  ${\bf A}({\bf r})$ remains the same as in vacuum, so the effect of plasmon modes may be ignored. Applying the nonequilibrium Green's functions formalism to the system of interacting electrons and photons one may derive the expressions for the radiated power, optical angular momentum and photon current. The relevant Green's functions are derived in earlier works \cite{27,31,32}. Here we merely reproduce some results essential for further analysis.

In general, the radiated energy flux is given by the expression:
\be
P=\oint{\bf S}\cdot d{\bf\Sigma}                                 \label{4}
\ee
where ${\bf S}$ is the Poynting vector and ${\bf\Sigma}$ is the surface vector. In the far field limit the surface is supposed to be a sphere centered on the molecule junction whose radius is much greater than the bridge length. The radiated energy could be computed using the expressions for the electron and photon Green's functions derived in earlier works \cite{31,32}. In the far field limit we arrive at the following approximation for the energy radiated in the direction of helix axis:
\begin{align}
P_{l}({\bf r})=&\frac{\tilde{\alpha}}{4\pi^3e^2c^2r^2}
\nn\\ &\times
\mathfrak{Re}\left(\int_{0}^{\infty}d\omega i\hbar^2\omega^2\int d\Sigma Tr\{({\bf U}-{\bf R}{\bf R})\Pi^{<}(\omega)\}\right)   \label{5}
\end{align}
Here, $\tilde{\alpha}=\ds\frac{e^2}{4\pi\epsilon_0\hbar c}$ is the fine structure constant, $\omega$ is the photon frequency, ${\bf U}$ is $3\times 3$ identity matrix, ${\bf R}=\ds\frac{{\bf r}}{r}$, ${\bf R}{\bf R}$ is the $3\times 3$ dyadic tensor and $\Pi^{<}(\omega)$ is the photon self-energy which originates from the electron-photon interactions. Using $d\Sigma=r dr d\Phi$ where $\Phi$ is the azimuthal angle in the $xy$ plane we get the result which does not depend on $r$: 
\be
P_l=\int_{0}^{\infty} d\omega\left(J_{+}(\omega)+J_{-}(\omega)\right)                   \label{6}
\ee
where
\be
J_{\pm}(\omega)=-\frac{\tilde{\alpha}\hbar^2\omega^2}{8\pi e^2c^2}\sum_{\alpha,\beta}\mathfrak{Im}\left(\Pi_{\pm}^{\alpha\beta}(\omega)\right).            \label{7}
\ee
Here, $\{\alpha,\beta\}=\{L,R\}$ indicate the electrodes.
It was shown that to the lowest order of electron-photon interaction, $\Pi^{<}(\omega)^{\alpha\beta}$ may be approximated as follows \cite{31}:
\begin{align}
\Pi ^{\alpha\beta <}_{\mu\nu}(\omega)=&\frac{-ie^2}{2\pi}\int_{-\infty}^{\infty} dE\left(f_{\alpha}(E)-f_{\beta}(E-\hbar\omega)\right)
\nn\\ &\times
n(\hbar\omega-\Delta\mu)Tr\{v^{\mu}\tilde{A}_{\alpha}(E)v^{\nu}\tilde{A}_{\beta}(E-\hbar\omega)\}.           \label{8}
\end{align}
In this expression, $f_{\alpha}(E)$ and $f_{\beta}(E-\hbar\omega)$ are Fermi functions for the electrodes with chemical potentials $\mu_{\alpha}$ and $\mu_{\beta}$, $n(\hbar\omega-\Delta\mu)$ is the Bose-Einstein distribution functions for photons, $\Delta\mu=\mu_{\alpha}-\mu_{\beta}$, $v^{\mu}$ are electron velocity components operators :
\be
v^{\mu}_{mn}=\frac{1}{\hbar}H_{mn}\left({\bf r}_m-{\bf r}_n\right).                       \label{9}
\ee
Matrices $\tilde{\bf A}_{\alpha}(E)$ are determined by the retarded ${\bf G}^{r}(E)$ and advanced ${\bf G}^{a}(E)$ Green's functons for electrons on the molecular bridge: $\tilde{\bf A}_{\alpha}(E)={\bf G}^{r}(E){\bf\Gamma}_{\alpha}{\bf G}^{a}(E)$ where the matrices ${\bf\Gamma}_{\alpha}$ describe the coupling of the molecular bridge to the electrodes.

Assuming that the couplings between the bridge site is sufficiently strong we keep both NN and NNN terms in the bridge Hamiltonian including $M$ sites. Also, we assume that matrix elements $H_{n-1,n}=H_{n+1,n}=t$ ($2\leq\leq n\leq M-1$) and $H_{n-2,n}=H_{n+2,n}=\gamma$ ($3\leq n\leq M-2$). Then we may write the following equation for $M\times M$ matrix ${\bf G}^{r}(E)$:
\begin{align}
&{(\bf G}^{r}(E))^{-1}=
\nn\\ 
&\left(\ba{cccccccc}{E-i\Gamma_L}&-t&-\gamma&0&0&0&...&0
\\
-t&E&-t&-\gamma&0&0&...&0
\\
-\gamma&-t&E&-t&-\gamma&0&...&0
\\
0&-\gamma&-t&E&-t&-\gamma&...&0
\\
...&...&...&...&...&...&...&...
\\
0&0&0&-\gamma&-t&E&-t&-\gamma
\\
0&0&0&0&-\gamma&-t&E&-t
\\
0&0&0&0&0&-\gamma&-t&{E-i\Gamma_R}
\\
\ea\right)                                \label{10}
\end{align}
Nonzero elements $M\times M$ matrices ${\bf \Gamma}_{\alpha}$ equal: $G^{11}_L=G_L$ and $\Gamma^{MM}_R=\Gamma_R$.

Using the expressions for ${\bf G}^{r}(E)$ and for ${\bf\Gamma}_{\alpha}$ it is easy to compute electron transmission $\tau(E)=Tr\{{\bf G}^{r}(E){\bf \Gamma}_{\alpha}{\bf G}^{a}(E){\bf\Gamma}_{\beta}\}$ as well as the photon self-energy $\Pi ^{\alpha\beta <}_{\mu\nu}(\omega)$. An example of the energy dependence of electron transmission is shown in Fig.1 (right panel). One sees that the transmission profile is not symmetrical with respect to $E=E_0$ as it occurs when NNN couplings are neglected. Further we show that this symmetry violation is a fundamentally important factor controlling the CP light emission.

 At low temperatures $\Pi^{<}_{\pm}(\omega)$ may be approximated as:
\be
\Pi^{<}_{\pm}(\omega)=-\frac{i e^2}{2\pi}\sum_{\alpha,\beta}\Theta(\Delta\mu-\hbar\omega)\int_{\mu_{\beta}+\hbar\omega}^{\mu_{\alpha}}dE X^{\alpha\beta}_{\pm}(E,\hbar\omega)                 \label{11}.
\ee 
In these expressions, $X^{\alpha\beta}_{\pm}=Tr\{v_{\pm}\tilde{A}_{\alpha}(E)v_{\mp}\tilde{A}_{\beta}(E-\hbar\omega)\}$, $v_{\pm}=v^x\pm iv^y$ and $\Theta(E)$ is the Heaviside step function. 

It is necessary to clarify the meaning of $J_{\pm}(\omega)$. Within the adopted model, charge carriers (electrons and holes) pushed by the bias voltage spiral in opposite directions along the helical bridge. Correspondingly, they pick up a counterclockwise/clockwise self rotation depending on the direction of their motion. The quantities $J_{\pm}(\omega)$  represent contributions to the energy of emitted light originating from optical transitions involving charge carriers (electrons/holes) rotating counterclockwise ($J_{+}(\omega)$) and clockwise ($J_{-}(\omega)$). As discussed below, this self-rotation of charge carriers causes the specifics of CP light emission. Therefore, the following analysis based on Eqs.(\ref{7}),(\ref{11}) is focused on the behavior of $J_{\pm}(\omega)$. 
 
\subsection{III.  Results and discussion.}

\begin{figure}[t] 
\begin{center}
 \includegraphics[width=4.2 cm,height=3.1cm]{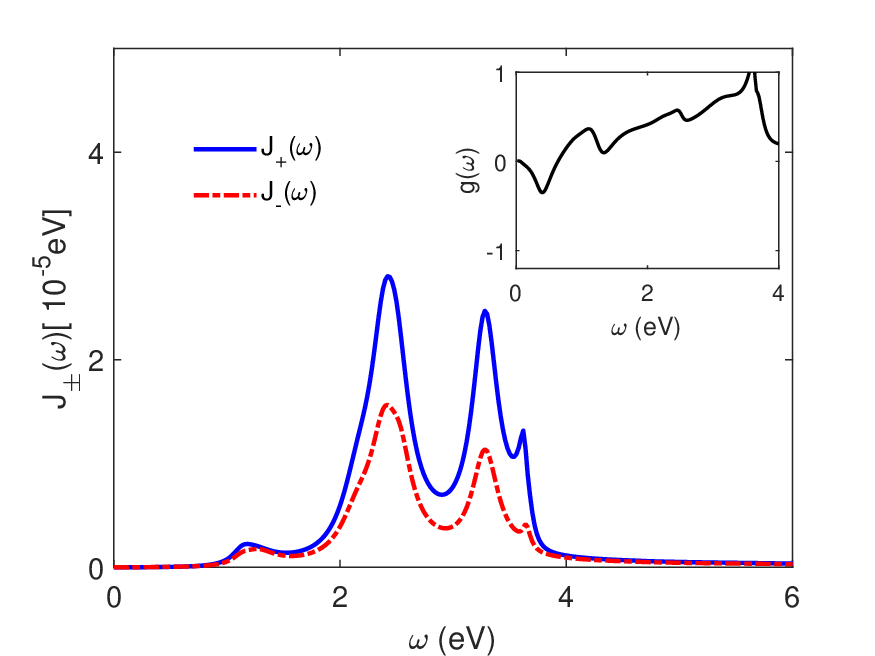}
\includegraphics[width=4.2 cm,height=3.1cm]{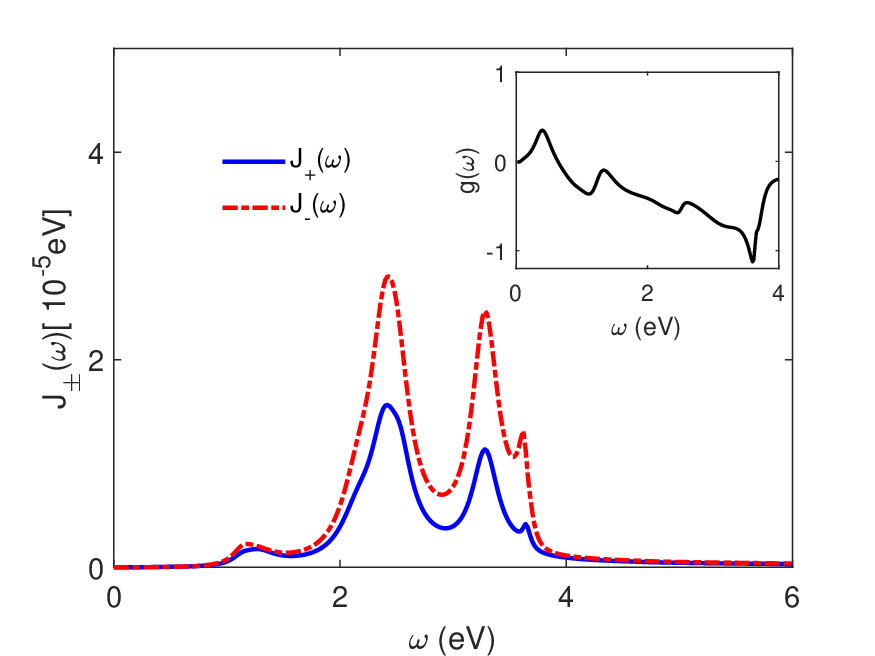}
\caption{$J_{\pm}(\omega)$ plotted at $M=6$, $E_0=0$, $t=1$eV, $\Gamma_L=\Gamma_R=0.2$eV $\gamma=0.4$eV, $r_g=0.5$nm assuming that $\mu_L>\mu_R$ (left) and $\mu_L<\mu_R$ (right)
}
\label{rateI}
\end{center}\end{figure}
First we analyze the CP light emission in a biased molecular junction assuming that the bias voltage is symmetrically distributed between the electrodes. We consider a large-bias limit which means that all bridge states $\{E_m\}$ are situated in the conduction window. Neglecting for a while NNN couplings between the bridge sites, we get energies $E_m$ symmetrically arranged around the equilibrium Fermi energy $E_F=E_0=0$, so the system possesses an electron-hole symmetry. For each permitted optical transition between two levels located above the Fermi energy ('electron' transitions) there exists a counter-part transition between the levels located below $E_F$ ('hole' transitions), and photons emitted in these transitions have the same energy. When the system is biased, electrons and holes traveling along a spiral pathway acquire  nonequilibrium orbital polarization and the longitudinal orbital angular momentum (OAM) is parallel to the direction of the charge carriers motion along the bridge axis. Thus the orbital-momentum locking discussed in earlier works \cite{12,40} is an inherent property of the considered system.
 
The  OAM may be transferred to emitted photons.
Photons of the same frequency emerging in 'electron' and 'hole' transitions get opposite  OAM for the relevant charge carriers spiral along the bridge in opposite directions. The longitudinal angular momentum transferred to the photons should be conserved.To ensure the conservation, light emitted forward and backward along $z$ axis as a result of identical optical transitions should have different handedness. For example, if $\mu_L>\mu_R$  and the current flows in the forward (z) direction through the molecular bridge of a shape shown in Fig.1 'electron' optical transitions produce left-handed CP light emitted backward and right-handed CP light emitted forward. The handedness of CP light emitted in the 'hole' transitions should be opposite. However, the electron-hole symmetry together with the symmetrical distribution of the bias voltage prevents the violation of the TRS. As a result, OAM transferred to the photons in the 'electron'  and 'hole' optical transitions counterbalance each other. 

The situation dramatically changes when the electron-hole symmetry is broken by NNN couplings which destroy the symmetrical arrangement of the bridge energy levels, as implied by the electron transmission shown in in Fig.1. In this case the optical transitions between 'electron' and 'hole' energy levels  are not balanced and $J_{+}(\omega)\neq J_{-}(\omega)$, as illustrated in Figs.2,3. The difference between $J_{+}(\omega)$ and $J_{-}(\omega)$ determines the OAM transferred by electrons on the bridge to the photons and may be characterized by the dissymmetry factor $g(\omega)$ defined as:
\be
g(\omega)=\frac{J_{+}(\omega)-J_{-}(\omega)}{(J_{+}(\omega)+J_{-}(\omega))/2}                       \label{12}
\ee

The destruction of electron-hole symmetry accompanied by the effect of the electric current flowing through the system destroys the TRS and opens up the possibility for the anomalous CP light emission. The left panel of Fig.2 is plotted assuming that $\mu_L>\mu_R$. For the chosen geometry of the helical bridge $J_{\pm}(\omega)$ are corresponding to the photons respectively emitted in 'electron' and 'hole' optical transitions along $z$ axis. In this case holes acquire an OAM directed along $z$ axis whereas OAM gained by electrons is oppositely directed (along $-z$). The contribution from 'hole'  transitions $J_{+}(\omega)$ exceeds the contribution from 'electron' transitions $J_{-}(\omega)$. Therefore, CP light emitted forward (backward) should be left (right) handed. The total OAM transferred to the emitted photons is proportional to the difference $(J_{+}\omega)-J_{-}(\omega))/\hbar\omega$ and directed along $z$ which agrees with the corresponding result of Ref.\cite{32}. When the bias voltage polarity is inverted ($\mu_L<\mu_R$), $J_{+}(\omega)$ is associated with 'electron' and $J_{-}(\omega)$ with 'hole' transitions. The OAM transferred to the photons becomes negative and the handedness of the light emitted forward and backward is reversed. This is shown in the right panel of Fig.2. 
\begin{figure}[t] 
\begin{center}
 \includegraphics[width=4.2 cm,height=3.1cm]{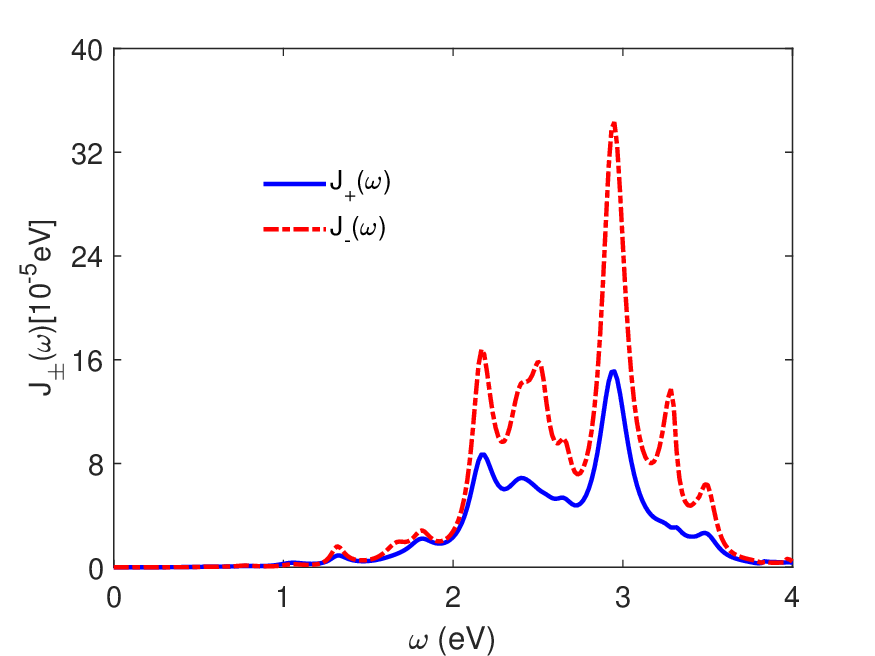}
\includegraphics[width=4.2 cm,height=3.1cm]{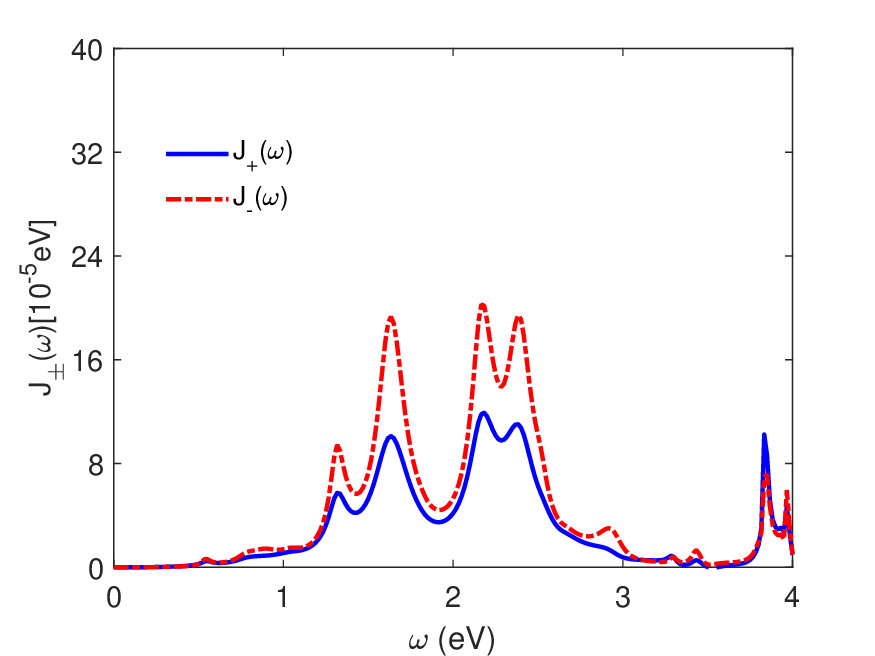}
\caption{ $J_{\pm}(\omega)$ plotted at $M=10$, $E_0=0$, $t=1$eV, $\Gamma_L=\Gamma_R=0.2$eV, $\gamma=0.4$eV assuming that $\mu_L<\mu_R$ at $r_g=0.5$nm , $\Delta\Phi=2\pi/5$ (left) and $r_g=1$nm, $\Delta\Phi=\pi/5$ (right). 
}
\label{rateI}
\end{center}\end{figure}
Note that $J_{\pm}(\omega)$ show several peaks reflecting the positions of the bridge energy levels and the difference between the two varies as the frequency changes. This means that the transferred to photons OAM is frequency dependent and the effect of nonreciprocal CP light emission may be stronger or weaker pronounced at different light frequencies.

The values taken by $J_{\pm}(\omega)$ noticeably increase when the bridge lengthens, as illustrated in Fig.3 where we present results for $M=10$. The effect of the varying bridge length depends on the bridge geometry. In two panels of this figure  we show results for $J_{\pm}$ computed for the same bridge length but for different scope of the azimuthal angle $\Phi_{n}$ indicating  positions of the bridge sites. The curves plotted at the left panel represent the case when the difference $\Delta\Phi$ between the adjacent bridge sites equals $2\pi/5$, and the bridge chain makes two turns about the helix axis whereas those plotted in the right panel are characterized by $\Delta\Phi=\pi/5$, so that the bridge makes a single turn. Comparing the displayed curves one may conjecture that increase in the number of turns of the helical bridge may noticeably strengthen the effect at the same bridge length and the same coupling strengths between the bridge sites.

\begin{figure}[t] 
\begin{center}
 \includegraphics[width=4.2cm,height=3.1cm]{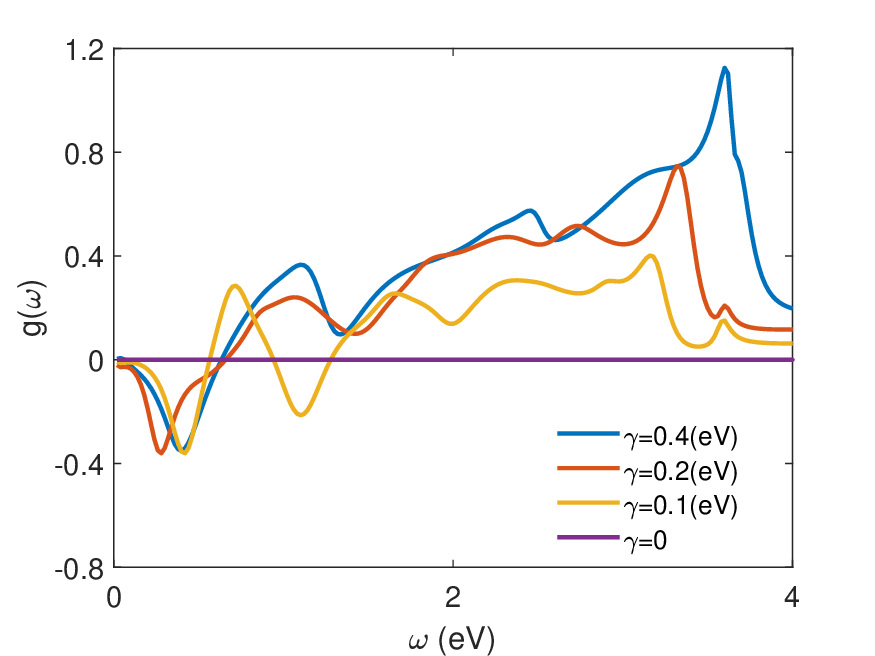}
\includegraphics[width=4.2cm,height=3.1cm]{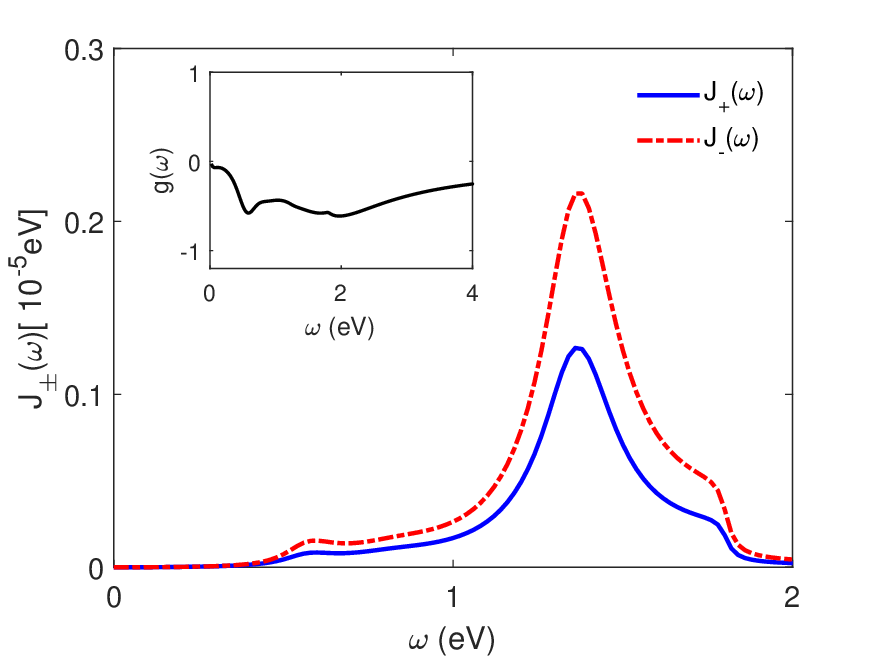}
\caption{ The dissymmetry factor $g(\omega)$ plotted at symmetrically distributed bias ($\mu_{L,R}=\pm \ds\frac{1}{2}|e|V$) at several values of NNN coupling strength $\gamma$ (left) and $J_{\pm}(\omega)$ plotted in the case of asymmetric bias ($\mu_L=0$, $\mu_R=|e|V$) (right) assuming that $M=6$, $E_0=0$, $t=1$eV,$\Gamma_L=\Gamma_R=0.2$eV, $r_g=0.5$nm, $\gamma=0$ (right). 
} 
\label{rateI}
\end{center}\end{figure}

The part played by NNN couplings in violation of TRS is further confirmed by the results displayed in the left panel of Fig.4. It  is shown that when the bias voltage is symmetrically distributed between the electrodes and NNN coupling between the bridge sites weakens, the difference between $J_{+}$ and $J_{-}$ diminishes, becoming zero when the electron-hole symmetry is restored. 

Another way to break TRS opens up when the symmetry of the bias voltage distribution is removed, so that some of the bridge energy levels fall out of the conduction window. In this case the electron-hole symmetry collapses and the system may emit CP light even when NNN interactions between bridge sites are negligible, as illustrated in the right panel of Fig.4.

\subsection {IV. Conclusions}

In summary, in the present work we used a simple model of an electrically driven single -molecule junction with a chiral linker shaped as a spiral chain to theoretically analyze properties of light emitted in optical transitions. The analysis was based on the nonequilibrium Green's functions formalism for interacting electrons and photons developed in previous works \cite{27,31,32}. It was shown that the adopted model simulates a system where TRS may be broken when the system is biased and the electron-hole states symmetry is removed due to sufficiently strong couplings  between the bridge sites so that at least NNN couplings should be taken into account.
We showed that the considered system may emit CP light whose handedness depends on both bias voltage polarity and the direction of propagation along the helical axis (forward/backward). 

The mechanism responsible for this anomalous electroluminescence is based on the transfer of the longitudinal orbital angular momentum from electrons to the photons emitted in optical transition and is not related to electromagnetic couplings and spin-orbit interactions. The angular momentum radiation controlled by the similar mechanism was analyzed in recent works \cite{27,32}. It is based on the orbital-momentum locking which appears in systems with specific electronic properties. The considered model is an example of such system. Other, more realistic systems of this kind, were previously studied \cite{12,41,42}. Nevertheless, the adopted model has an advantage of catching the essential physics without carrying on cumbersome band structure calculations. We believe that the results of the present research may contribute to further understanding of optical properties of single molecule junctions with chiral linkers.

\subsection{Declaration of competing interest}

Authors declare that they have no competing financial interests or personal relationships which could influence the work reported in this paper.

\subsection{Data availability statement}

Data sharing is not applicable as no data are created in this study.

\subsection{Acknowledgments}

The present work was supported by the U.S National Science Foundation (DMR-PREM 2122102).

\end{document}